\newif\iftimes
\timestrue

\documentclass[submission,copyright,creativecommons]{eptcs}
\providecommand{\event}{DSL 2011} 

\iftimes
\usepackage{t1enc,times}
\font\cour=pcrb8r at 10pt
\else
\usepackage{t1enc}        
\fi

\usepackage{alltt}
\usepackage{amssymb}
\usepackage[pdftex]{graphicx}

\makeatletter
\iftimes
\def\verbatim@font{\cour}

\def\tt{\cour}
\else
\def\verbatim@font{\normalfont\ttfamily\small}

\fi

\makeatother
\def\itbf{\itshape\bfseries} 

\usepackage{wrapfig,float}

\input supp-pdf
\def\mnd{\par\medskip\noindent}

\newcommand\ovl[1]{\overline#1}

\def\lam#1{\char"5C{#1}->}

\title{Specific ``scientific'' data structures, and their processing}
\author{Jerzy Karczmarczuk
\institute{Dept. of Computer Science, University of Caen, France}
\email{jerzy.karczmarczuk@unicaen.fr}
}
\def\titlerunning{Scientific data structures}
\def\authorrunning{J. Karczmarczuk}
\begin{document}
\maketitle

\begin{abstract}
\noindent
Programming physicists use, as all programmers, arrays, lists, tuples, records, etc., and this requires some change in their thought patterns while converting their formulae into some code, since the ``data structures'' operated upon, while elaborating some theory and its consequences, are rather: power series and Padé approximants, differential forms and other instances of differential algebras, functionals (for the variational calculus), trajectories (solutions of differential equations), Young diagrams and Feynman graphs, etc. Such data is often used in a [semi-]numerical setting, not necessarily ``symbolic'', appropriate for the computer algebra packages. Modules adapted to such data may be ``just libraries'', but often they become specific, embedded sub-languages, typically mapped into object-oriented frameworks, with overloaded mathematical operations. Here we present a functional approach to this philosophy. We show how the usage of Haskell datatypes  and~-- fundamental for our tutorial~-- the application of lazy evaluation makes it possible to operate upon such data (in particular: the ``infinite'' sequences) in a natural and comfortable manner.
\end{abstract}

\vspace{0.3cm}
\rightline{\em The fox knows many little things; the hedgehog one big thing.}
\rightline{Archilochus, 680 BC -- 645 BC}

\bigskip

\section{Introduction}

\noindent
The most ubiquitous boiler-plate code element in scientific computation are iterative loops. The term: ``scientific computation'' means here applied mathematics, including algebra, analysis and geometry. We shall not speak about ``symbolic'' computation -- the processing of syntactic structures containing symbolic indeterminates -- but about numerical and ``semi-numerical'' (using the Knuth \cite{KNUTH} terminology)
\begin{wrapfigure}[10]{r}{3.3cm}
  \centering
  \vspace{-0.4cm}
  \includegraphics[width=3.3cm,keepaspectratio]{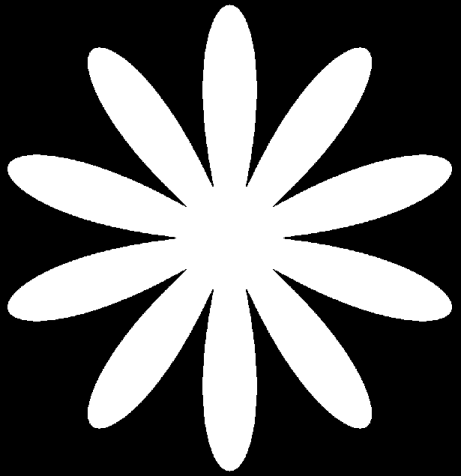}
  \vspace{-0.7cm}
  \caption{An algorithmic flower}
\end{wrapfigure}
calculi. 
Semi-numerical means that the processed data are mainly numeric, but usually composite and structured, subject to specific global mathematical operations, e.g.~the multiplication of powers series represented as sequences of coefficients, the contraction of tensors, etc.

Some of these structures are relatively primitive. If $x$ is an appropriately normalized horizontal gradient picture -- a two-dimensional array of numbers representing gray pixels, and $y$ -- its transpose (a vertical gradient), the image at the right is obtained by one simple algebraic expression: $x^2+y^2 < 0.5-0.45\cdot \cos(10\cdot \mathrm{atan2}(y,x))$.

The {\bf for} loop for the pixel coordinates is implicit and hidden, and since in the domain of scientific computing, simulation, etc., the loops are omnipresent, this vectorization mechanism is the main source of the success of such languages as Matlab. But in images pixels are independent. In other data structures -- such as the mentioned above power series -- the manipulation of objects combines many elements in a non-trivial manner. In others, such as sequences representing the numerical solutions of differential equations, their construction is incremental, takes into account many existing data pieces in order to add more. This cannot be easily vectorized.

This text advocates the use of co-recursive, lazy data structures (mainly 1-dimensional sequences) processed through functional algorithms implemented in Haskell \cite{HASKE}. We will show several {\em heterogeneous examples} of the same, uniform approach: how to treat these objects as mathematical expressions, how to build complicated algorithms upon them, and how to adapt some delicate computational problems to this sort of algorithmization. The understanding of a simple Haskell code is required from the reader, but no real experience with lazy algorithms is assumed.

\subsection{Simple co-recursive sequences}
The laziness of Haskell is often introduced through simple examples of non-terminating, extrapolating recursive functions, which generate ``infinite'' lists, such as the definition of the sequence of all positive integers: {\tt integs}, predefined as {\tt [1 ..] = [1, 2, 3, 4, ...]}
\begin{alltt}
integs = fromN 1 where fromN m = m : fromN (m+1)
\end{alltt}
We shall be more interested by {\em recursive data} rather than recursive functions. The case above may be defined as
\begin{alltt}
integs = 1 : (integs + ones) where ones = 1 : ones 
\end{alltt}
(From now on, the operator {\tt (+)} acting on sequences, it should be read as the overloaded, element-wise addition, equivalent to {\tt zipWith (+)}, or similar. Our sequences may not be standard lists, but they should be Functors (possessing the mapping functional), and they should be ``zippable''). The basic programming pattern is the definition of a piece of data in terms of itself, but ``protected'' by a known prefix. So, the list {\tt ones} begins with 1, and that permits the program to establish the value of its second elements, which is $\ldots\!$ 1, etc. The second element of {\tt integs} is equal to $1+1=2$, which makes it possible to compute the third element. The essence of the algorithm consists in the sequential traversal of the sequence; no element can be created before its prefix. This is a similar definition of the Fibonacci sequence {\tt [0, 1, 1, 2, 3, 5, 8, 13, 21, 34, ...]}:
\begin{alltt}
fibs = 0 : ftail where ftail = 1 : fibs + ftail
\end{alltt}
We observe that such constructions are too often presented as pedagogical curiosities, since this ``borrowing from the future'', {\em extrapolating} recursion, cannot be easily implemented in older classical languages. We want to advocate their usage as a {\bf genuine, practical programming methodology}, or -- perhaps -- a specific language. We have observed that the obvious activity of physicists, engineers, etc.\ of ``doing mathematics on a computer'' too often results in an {\em abuse} of computer algebra packages because of the confusion between mathematics and the processing of formal expressions. The paradox is that very often this manipulation of formulae with indeterminates produces the results unwieldy and unreadable, which are used {\em only} to generate numerical codes in, say, Fortran, not for insight.

We believe that a more rational approach would be to implement mathematical computational structures in a more direct way, and the presented philosophy may be a small step in this direction.

\section{Sound generators and transducers}
Perhaps it would be interesting to start not with ``just'' mathematical sequences, but with something having a practical physical meaning. A sound~\cite{SOUND} is for us just a sequence -- the discrete amplitude as a function of time, generated by algorithmic methods~\cite{PADL05}. For technical reasons we have written our package in Clean~\cite{CLEAN}, but here we present some functions coded in Haskell. The simplest, monochromatic acoustic stream is a sinusoid which could be created as {\tt y = map (\lam{n}sin(h*n)) [0 ..]}, but this is not natural: real sound generators (hardware or human) do not count the time (in this context). 
\begin{wrapfigure}[8]{l}{6.0cm}
  \centering
 \includegraphics[width=6.0cm,keepaspectratio]{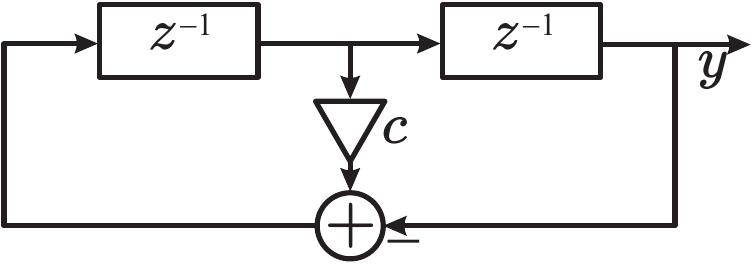}
 \vspace{-0.70cm}
\caption{Sine generator}
\end{wrapfigure}
Many synthesizers use the recurrence formula
\begin{equation}
\sin(n h) = 2 \cos(h) \cdot \sin((n-1)h) - \sin((n-2)h)
\end{equation}
which can be depicted as the diagram at the left. The block $z^{-1}$ is a delay element, and the triangle depicts a multiplication by a scalar. This, and many other similar recurrences may be easily constructed as co-recursive streams. The recipe is straightforward, the full sequence is given by
\begin{alltt}
y = sin h : ((2*cos h*>yr) - (0:yr))
\end{alltt}
where the subtraction has been overloaded for lists, and {\tt x *> l = map (x*) l} is another useful universal operator which multiplies sequences by scalars. The conversion of recurrence formula into streams is the main technique covered by this tutorial.

\begin{wrapfigure}[7]{r}{6.3cm}
  \centering
 \vspace{-0.80cm}
 \includegraphics[width=6.3cm,keepaspectratio]{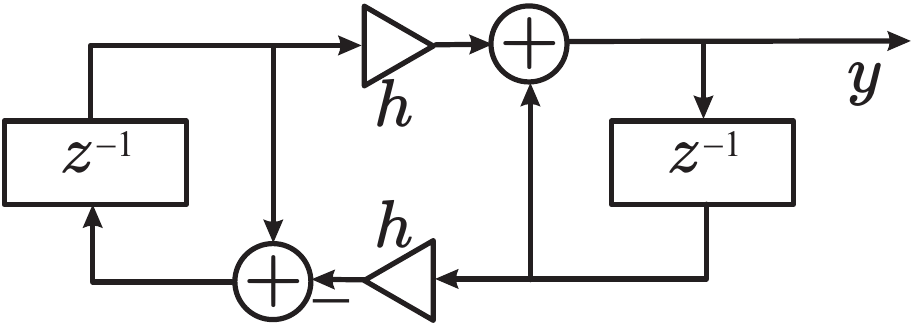}
 \vspace{-0.70cm}
\caption{Another oscillator}
\end{wrapfigure}
\begin{figure}[H]
\centering
 \includegraphics[width=16cm,keepaspectratio]{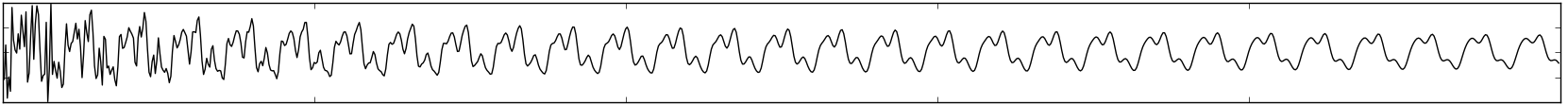}
 \vspace{-0.70cm}
 \caption{Sound of a plucked string}
 \label{kswav}
\end{figure}

There is another popular harmonic oscillator implementation, usable (sufficiently precise) for a very small $h$, based on the numerical solution of the appropriate differential equation (with the frequency $\omega=1$): $y'' = - y$ by a modified and stable Euler method, with $v$ denoting $y'$: $y_{n+1} = y_n + h\cdot v_n$;  $v_{n+1} = v_n - h\cdot y_{n+1}$. The ``lazy conversion'' in this case is also easy:
\begin{alltt}
y=0:w  where \{w=y+h*>u ; u=1:u-h*>w\}
\end{alltt} 
Observe that the ``attenuation'' {\tt h*>u} etc.\ does not determine the amplitude, but the frequency of the signal. If we replace these terms by a product with another, slowly oscillating stream: {\tt slowOscil*u}, etc., we will get a {\em vibrato} effect. More complicated ``soft musical instruments'' may be based on similar principles.\newline
\begin{wrapfigure}[6]{l}{6.0cm}
  \centering
 \vspace{-0.60cm}
 \includegraphics[width=6.0cm,keepaspectratio]{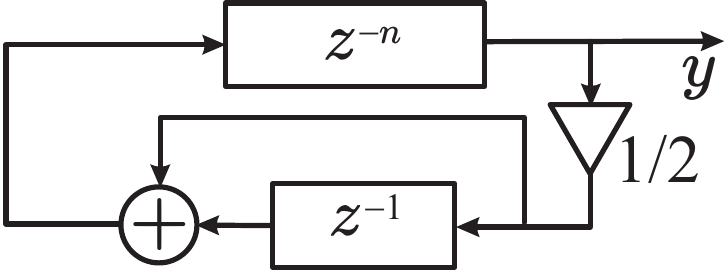}
 \vspace{-0.70cm}
\caption{K-S string}
\end{wrapfigure}
A model of a plucked string proposed by Karplus and Strong~\cite{KASTRO}, whose generalizations are common ingredients of many string but also wind instruments, contains a delay line representing the wave propagating medium. Its length determines the basic frequency of the string. 
For us, this will be a finite list, initialized by anything, e.g.~by a random noise. This list is the prefix of the infinite stream, whose assembly {\em filters} the sequence, eliminating the high frequencies as shown on the Fig.~(\ref{kswav}). The code is quite trivial
\begin{alltt}
y = prefx ++ 0.5*>(y + (0:y))
  where prefx = take n random_stream
\end{alltt}
and the result is a fairly realistic sound, similar to the guitar, or koto, or other classical string instruments. The details depend on the initial excitation, and on the filter used. Our approach permits the construction  of complicated filters and other transducers (reverberators, {\em vibrato} contraptions, etc.~\cite{JOSWA}) by an {\em extremely} compact code.

\subsection{Filters}
We have already seen a simple FIR ({\em Finite Impulse Response}) filter -- the mean of two neighbouring elements of the input stream. This weakens the higher frequencies and attenuates the signal. The laziness is not needed for its implementation. But a general linear (IIR: {\em Infinite Impulse Response}) filter is a transformation  $x \to y$ described by
\begin{equation}
y_n = \sum_{k=0}^m b_k x_{n-k} + \sum_{k=1}^p a_k y_{n-k}\,,
\end{equation}
and it may exploit the co-recursion more actively. One useful category of filters is the {\em all-pass}, whose simple instance is: $v_n = x_n -b\cdot v_{n-M}; \quad y_n = b\cdot v_n + v_{n-M}$. \newline
\begin{wrapfigure}[7]{l}{9.0cm}
  \centering
 \vspace{-0.60cm}
 \includegraphics[width=9.0cm,keepaspectratio]{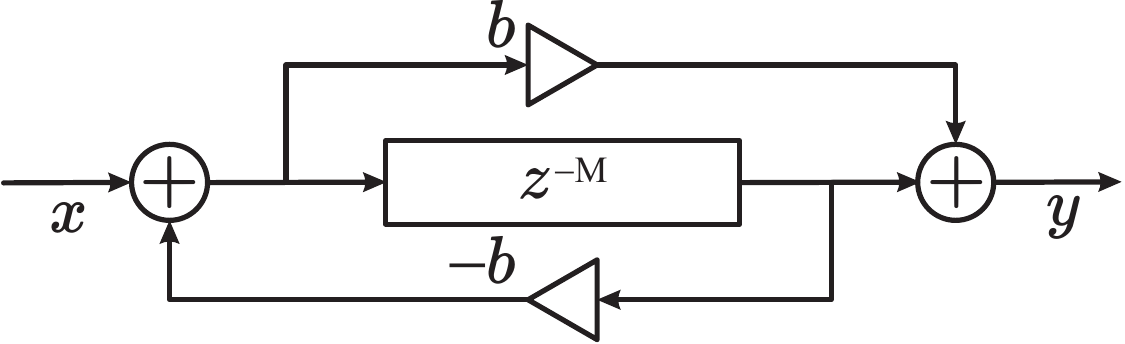}
 \vspace{-0.80cm}
\caption{An all-pass filter}
\label{alpas}
\end{wrapfigure}
Its coding follows our standard co-recursive pattern. We can define it as the following function.
\newline
{\tt allpass m b x = b*>v + d where \newline
\hbox to 1em{} d = delay m v\newline
\hbox to 1em{} v = x - b*>d\newline
}
The all-pass filters do not attenuate particular frequencies, but introduce the {\em dispersion} of sound, as if the waves with different frequencies had different velocities. They are useful for the percussion instruments, for such effects as the reverberation, and for coding fractional delays, needed for the fine tuning of the musical note frequency.

The main advantage of the co-recursive approach is the compactness of the code, but also its modularity. But we should pass to more formal, mathematical sequences.


\section{Power series}
Most approximations in mechanics, etc.\ are based on perturbational calculus, which yields the result as a power series depending on a small parameter, say, $x$: $U=u_0 + u_1 x + u_2 x^2 + \cdots{} + u_n x^n + \cdots{}$ Almost always a relatively small number of coefficients $u_k$ is computable, and the convergence of the series is a secondary issue, in most interesting cases the series are anyway asymptotic only. In the formal processing of series the variable $x$ remains indeterminate and is never really used. We will be interested only in the set of coefficients $[u_0, u_1, u_2, \ldots]$~\cite{LAZYSEM,MCILROY}.

Adding and subtracting series is easy, element-wise. The multiplication is a convolution, which requires loops, and a fair amount of ``administrative'' code, in order to ensure the correct treatment of truncations. But we don't truncate anything\ldots{} Let's represent the series through their first element and its tail: $U = u_0 + x \ovl{U}$, where $\ovl{U} = u_1 + u_2 x + \cdots$~\cite{LAZYSEM}. We can immediately write
\begin{equation}
U \cdot V = (u_0 + x \ovl{U})\cdot (v_0 +x \ovl{V}) = u_0\cdot v_0 + x(u_0 \ovl{V} + \ovl{U}\cdot V)
\end{equation}
One doesn't need to be an expert in Haskell in order to code the overloaded multiplication:
\begin{alltt}
(u0 : uq) * v@(v0 : vq) = u0*v0 : (u0*>vq + uq*v)
\end{alltt}
The definition is recursive, but since the prefix (the first element of the result) is available immediately, the {\em second} element is computable as well. This is a sound co-recursion, with finite progress. The division seems more intricate, and its ``classical'' code takes a half of a typical textbook page, or more. But, if $W=U/V$, then $U=W\cdot V$, and we see that in $U = u_0 + x \ovl{U} = v_0 w_0 + x(w_0 \ovl{V} + \ovl{W}\cdot V)$, the components of $W$ are computable:
\begin{alltt}
(u0 : uq) / v@(v0 : vq) = 
  let w0 = u0/v0
  in  w0 : (uq - w0*>vq) / v
\end{alltt}
and here the laziness of the language is much more ``aggressive'', transposing codes written in this style to a strict, imperative language is really difficult.

The classical elementary functions need just two auxiliary functions: the differentiation and the integration of series:
\begin{alltt}
sdiff (_ : uq) = zipWith (*) uq [1..]
sint cnst u = cnst : zipWith (/) u [1..]
\end{alltt}
If $W = e^U$, then its derivative $W' = e^U\cdot U'$, or $W = \int W\cdot U'$. We have thus another 1-liner:
\begin{alltt}
exp u@(u0 : _) = w where w = sint (exp u0) (sdiff u * w)
\end{alltt}
Exactly the same procedure can be used for the square root: if $W=\sqrt{U}$, then $W' = U'/(2\sqrt{U})$, or $W = \sqrt{u_0} + \int{U'/(2 W)}$, or for the arbitrary power. If $W = U^a$, then $W' = a U^{a-1} U'$, or $W = u_0^a + \int{U'\cdot W/U}$.

A recurrent problem in computation with series is that often the theory gives us a perturbation expansion in the wrong direction (e.g.\ we get a power series expressing some thermodynamic potential through pressure, but we want to compute the pressure!) Given a series without the free term (this is important), say
\begin{equation}
z = t + v_2 t^2 + v_3 t^3 + \cdots,
\label{invser}
\end{equation}
we want to construct its functional reversal, the series $t=W(z)$, such that $z + w_2 z^2 + w_3 z^3\cdots = t$. The algorithms are known since the times of Lagrange, but their coding is quite heavy. But we can reduce the algorithmization to the {\em composition of series}, which produces a very short code. We begin thus with another question, how to construct $W(x) = U(V(x))$, where $V(0)=0$, otherwise we would have to compute an infinite numerical sum, in order to get $w_0$. The co-recursive coding in this case is just an infinite Horner scheme
\begin{eqnarray}
U(v) &=& u_0 + V\cdot(u_1 + V\cdot (u_2 + V\cdot (u_3 + \cdots)))\\
 &=& u_0 + x(v_1 + x v_2 + \cdots)\cdot(u_1 + x(v_1 + x v_2 + \cdots)\cdot(u_2 + x(\cdots)\cdot(u_3 + \cdots)))\,.
\end{eqnarray}
This expression contains just the multiplication of series in a recursive, obvious pattern. The code is:
\begin{alltt}
scomp u (_: v) = cmv u where
  cmv (u0 : uq) = u0 : v*cmv uq
\end{alltt}
We may return to (\ref{invser}). We cannot code directly $t = z - v_2 t^2 - v_3 t^3 - \cdots$, since this is not properly co-recursive (try!) and will not terminate. But if we introduce $p$ such that $t=zp$, the ``miracle happens'', the expression $p = 1 - z\cdot p^2 [v_2 + v_3 t + v_4 t^2 + \cdots]$ may be coded :
\begin{alltt}
revser (_ : _ : vb) = t where
  t = 0 : p
  p = 1 : (-p*p*scomp vb t)
\end{alltt}
The lesson is the following: lazy algorithms demand sometimes an intelligent preprocessing of the available data, in order to construct sound co-recursive definitions. Such techniques are not obvious, and require some experience. This is not the question of knowing a library, or becoming fluent in Haskell. The user should master {\em a specific language semantics}, in order to  formulate his algorithmization before writing the code.

We will try to solve a {\em singular} differential equation, e.g.\ the modified Bessel equation:
\begin{equation}
x^2 w'' + w' + \frac{1}{4} w = 0\,.
\end{equation}
We cannot integrate twice the $w''$ term because of the singularity. (We could write an extension for the Laurent or the Puiseux series, but here we don't want to). We know that a regular solution exists, and it suffices to integrate once $w' = -x^2w'' - \frac{1}{4} w$ with $w_0=1$. This equation {\bf is} the solution:
\begin{alltt}
w = sint 1 (-(1\%4)*>w - (0:0:sdif (sdif w)))
\end{alltt}
yielding
\begin{equation}
w = [1, -\frac{1}{4}, \frac{1}{32}, -\frac{3}{128}, \frac{75}{2048}, -\frac{735}{8192}, \frac{19845}{65536},\ldots]
\end{equation}
Other strategies are also possible, but require more human work.

At the beginning we have used simple lists to represent series (and other sequences), but this is not the best choice. First, lists are commonly used for other purposes, and it is better not to overload them with all possible meanings, Haskell offers the algebraic data types, with which it is easier to define the numerical type classes. In fact, our true series package uses also a specific datatype
\begin{alltt}
data  Series a = Zs | !a :* Series a
\end{alltt}
where {\tt Zs} is a compact zero series, avoiding the usage of an infinite list when not needed. The definition of the arithmetic operations is of course almost the same as before.

\subsection{Partitions}
Another example is a generating function for the number of partitions of an integer. One of possible representations is an infinite product 
\begin{equation}
Z(x) = \prod_{n=1}^\infty \frac{1}{1-x^n}\,.
\end{equation}
which seems hardly usable. It may be reformulated as a ``runaway'', open recurrence
\begin{equation}
Z(x) = Z_1(x) \qquad \mathrm{where} \qquad Z_m(x) = \frac{1}{1-x^m} Z_{m+1}(x)\,.
\end{equation}
If we regard at $Z_m(x)$ as a series in $x$, the sequence of its coefficients starts with 1 followed by $m$ zeros. We can rephrase the equation above as: $Z_m = Z_{m+1} + x^m Z_m$. Notice that an attempt to express $Z_{m+1}$ through $Z_m$, the ``normal'' recursion, would be utterly silly. We need ``just'' $Z_1$\ldots{}

We define a family of auxiliary sequences $B_m$ such that $Z_m = 1 + x^m B_m$. It fulfils a sound co-recursive scheme
\begin{equation}
B_m(x) = 1 + x \left(B_{m+1}(x) + x^{m-1}B_m(x)\right)\,.
\end{equation}
The program:
\begin{alltt}
partgen = 1 :* b 1 where
  b m = p where p = 1 :* b (m+1) + (0:*1)^(m-1)*p
\end{alltt}
yields [1, 1, 2, 3, 5, 7, 11, 15, 22, 30, 42, 56, 77, 101,135, 176, 231, \ldots] without difficulties. We don't even need the series algebra, we have only used the integer power of a primitive series {\verb|(0:*1)^m|}, which can be coded as 1 prefixed by $m$ zeros.


\section{Differential expressions}
Everybody needs derivatives. They are not only delivered as some wanted final results, but play also {\em active} role: they are used in differential recurrences which define some functions, in the determination of the propagation of errors in processing data with uncertainties, etc. Sometimes simple numerical approximations (differential quotients) suffice, but in general such procedures are unstable, not suited for higher-order derivatives, and if iterated, the quality of approximation is difficult to estimate. The ``symbolic'', formal differentiation dominates, and the usage of computer algebra packages is fairly common, even if what we really need are numbers -- derivatives at a given value of the differentiation variable. (For simplicity we will discuss here the 1-dimensional case.)

We will show how to enlarge the domain of ``normal'' arithmetic expressions which may depend on {\em one} particular object called the ``differentiation variable'', whose name is irrelevant~\cite{KHOSC}. The standard algebra of expressions, with the 4 operations, elementary functions, etc., becomes a {\em differential algebra} -- a domain which contains also the {\itbf derivation}: $e \to e'$, which is a linear ($(e+f)' = e' + f'$) operator, obeying the Leibniz rule: $(e\cdot f)' = e'\cdot f + e \cdot f'$. This rule, and the existence of the division suffice to prove that $(e/f)' = (e' f - e f')/f^2$, and all the differentiation properties of standard functions. But the elements of such an algebra cannot be ``just numbers'', it is easy to prove that such an algebra is trivial, that $e'=0$ for all $e$.

We will implement the expressions belonging to a non-trivial algebra extensionally, as sequences whose head is the ``normal'' expression containing constants, the value of our ``variable'', etc., and the remaining elements are derivatives. The programs will manipulate lazy lists $[e,e', e'', e^{(3)}, \ldots]$. The derivatives are not algorithmically computed from a structural form representing the expression (the program doesn't keep it), but are physically present inside. The value of the first derivative of such expression is its second element. Since in general the derivative is algebraically independent of the original expression, the sequence is potentially infinite, only in the case of polynomials the sequence terminates with zeros.

\subsection{Arithmetic}
Of course, we have to start with something. Constants, such as $\pi$ are represented as $[3.14159\ldots, 0, 0,\ldots]$, and the differentiation variable whose value is, say, $2.5$, is $[2.5,1,0,0,\ldots]$. There is a resemblance between this domain and the (Taylor) power series, but they are mathematically different. In order to optimize the structure of these generalized expressions, we shall not use normal lists, but specific algebraic data structures:
\begin{alltt}
data Dif a = Cd !a | !a :> Dif a
\end{alltt}
where the first variant denotes a constant followed by zeros (no need to keep the full infinite tower, although this might be produced while processing these data), and the recursive case is equivalent to a list. The ``variable'' will be coded as {\tt 2.5 :> Cd 1}, or as {\tt 2.5 :> 1}, since the lifting from number to constants is automatic, with the instance {\tt fromInteger c = Cd (fromInteger c)}. The arithmetic is defined below, the presentation is simplified. The derivatives ``compute themselves'' as parts of the manipulated expressions.
\begin{alltt}
Cd x + Cd y = Cd (x+y)    -- {\em Trivialities}
Cd x + (y:>yq) = (x+y):>yq
(y:>yq) + Cd x = (x+y):>yq
(x:>xq)+(y:>yq) = x+y :> xq+yq

c*>Cd x = Cd (c*x)
c*>(x:>xq) = (c*x):>(c*>xq)
Cd x * p = x*>p -- {\em and symmetrically}
xs@(x:>xq) * ys@(y:>yq) = (x*y) :> (xs*yq+xq*ys)
sqr xs@(x:>xq) = x*x :> (2*>(xq*xs))   -- {\em Square. Slightly optimized}
\end{alltt}
The remaining functions are also straightforward, it suffices to transpose some school formulae into code, with a minimum of discipline. The co-recursion follows similar patterns as in the case of series.
\begin{alltt}
recip xs@(x:>xq) = ip where
  ip = recip x :> (-xq*sqr ip)
xs@(x:>xq)/ys@(y:>yq) 
  | x==0 && y==0 = xq/yq
  | otherwise = w where w = x/y :> xq/ys - w*yq/ys

exp xs@(x:>xq) = w where w = exp x :> xq*w
log xs@(x:>xq) = log x :> (xq/xs)
sqrt xs@(x:>xq) = w where w = sqrt x :> ((1/2) *> (xq/w))
sin xs@(x:>xq) = sin x :> xq*cos xs
cos xs@(x:>xq) = cos x :> (-xq*sin xs)
atan xs@(x:>xq) = atan x :> xq/(1+sqr xs)
asin xs@(x:>xq) = asin x :> xq/sqrt(1-sqr xs)
\end{alltt}
The division is redundant, it may be retrieved from {\tt recip}. We have included a variant which applies automatically the de l'Hôpital rule, but this is not a universal choice !

\subsection{Some applications}
This was a school exercise, real problems begin {\em now}, since the system is full of traps. The instantiation of the Leibniz rule, when iterated, may easily generate an exponential complexity for the higher-order derivatives. The life of a lazy programmer is very laborious, as the memory may easily clog with unevaluated thunks, if not surveyed. The evaluation of the expression {\tt sin x*exp(-x)} with {\tt x} being the differentiation variable, may generate, say, 24 derivatives, and the memory saturates. But would we, intelligent humans, apply blindly the Leibniz rule? We note immediately that the sine and cosine generate similar, alternating terms, and the function
\begin{alltt}
exsn x@(x0:>_) = p where
  ex=exp(-x0)
  p = sin x0*ex :> q - p
  q = cos x0*ex :> (-p - q)
\end{alltt}
produces a very large number of elements in a very short time.

The next example is the computation of the Maclaurin series for the Lambert function~\cite{CORLEKN} $W(z)$, defined by an implicit equation
\begin{equation}
W(z)\cdot e^{W(z)} = z\,.
\label{lamber}
\end{equation}
It is useful in combinatorics, and also in theoretical physics: solutions of several differential equations may be expressed through this function. The differentiation of (\ref{lamber}) gives
\begin{equation}
\frac{dz}{dW} = e^W(1+W) \qquad \left(=\frac{z}{W}(1+W) \quad \mathrm{for}\quad z \ne 0\right)\,,
\end{equation}
whose inversion
\begin{equation}
\frac{dW}{dz}= \frac{e^{-W}}{1+W} \qquad \left(\frac{W}{z} \frac{1}{1+W}\right)
\end{equation}
may be directly coded, taking into account that $W(0)=0$:
\begin{alltt}
w = 0.0 :> (exp (-w)/(1.0+w))
\end{alltt}
This gives the asymptotic series 0.0, 1.0, 2.0, 9.0, 64.0, \ldots, $(-n)^{n-1}$, \ldots{} , which is of course known, but the idea may be easily used in other expressions.


\subsection{The WKB asymptotic expansion}
\def\ve{\varepsilon}
The last example is a procedure useful e.g., in quasi-classical approximations in quantum physics, in the scattering theory, etc.~\cite{BENDER}, the Wentzel-Kramers-Brillouin scheme. We shall just sketch the problem and its solution, in order to show how the co-recursion avoids a very, very intricate coding gymnastics. This example should be understood, not necessarily tested. We begin with a simple, but singular differential equation for $y(x)$, parameterized by a very small number $\ve$:
\begin{equation}
\ve^2 y'' = Q(x) y\,.
\label{wkbeq}
\end{equation}
This is nothing exotic, the equation describes a (generalized) wave in space, with the frequency going to infinity (or the Planck constant going to zero). The function $Q(x)$ is usually simple (for the useful Airy function $Q(x)=x$), and we want to develop $y$ as a series in $\ve$. But $\ve=0$ is an essential singularity, which degenerates the equation (\ref{wkbeq}). Applied mathematics shows many examples of functions which go to zero when their argument goes to zero, but no power series can be proposed in its neighbourhood, e.g., $f(x) = \exp(-1/x)$. This is the case here\ldots{}

Within the WKB formalism, the proposed solution for $y$ is represented as (we neglect the normalization):
\begin{equation}
y \propto \exp\left(\frac{1}{\ve} \sum_{n=0} \ve^n S_n(x)\right)\,.
\end{equation}
Inserting that into the equation (\ref{wkbeq}) gives a chain of entangled recurrences for $S_n(x)$. The lowest approximation is $S'_0(x) = \pm\sqrt{Q(x)}$, which should be explicitly integrated, this will not be discussed. For the remaining terms it is natural to separate the even and the odd terms in the powers of $\ve$:
\begin{equation}
y=\exp\left(\frac{1}{\ve}S_0(x) + U(x;\ve^2) + \ve V(x;\ve^2)\right)\,.
\end{equation}
It is easy to see that injecting this formula into (\ref{wkbeq}) generates the following recurrences:
\begin{equation}
U' =\frac{-1}{2}\frac{S''_0 + \ve^2 V''}{S_0' + \ve^2 V'}\,, \qquad
V'= \frac{-1}{2S_0'}\left(U'^2 + U'' + \ve^2 V'^2\right)\,,
\end{equation}
which can be partly solved: $e^U = 1/\sqrt{S_0'+\ve^2V'}$, but it is not clear whether this strategy is sound, since $V'$ demands the value of $U''$, which requires $V''$ and $V^{(3)}$, etc. The classical algorithmization of the WKB approach is a nightmare\ldots{} But $U$ and $V$ are series in $\ve^2$ (not in $\ve$, and absolutely not in $x$!), and higher derivatives of $U$ and $V$ appear only in higher order terms in $\ve$. The co-recursive formulation of the process becomes effective. The program is just the transcription of the formulae above:
\begin{alltt}
s0'=\ldots{}{\em (depends on Q. The solution is for one value of x.)} :: (Dif Double)

infixl 6 +:
(a0 :* aq) +: z = a0 :* (aq+z)  --{\em auxiliary}

u = -0.5*>log(s0' :* v')
u' = fmap diff u
v' = (-0.5/s0')*>(u'*u' + fmap diff u' +: v'*v')
\end{alltt}
Now {\tt u', v'} are series whose coefficients are differential chains. From this chains we need only the principal values, not the derivatives, so we calculate {\tt fmap dmain u'} and {\tt vv' = fmap dmain v'}, where {\tt dmain (x:>\_) = x}. The only difficulty is the correct typing discipline, and ordering of the hierarchical structures: what kind of sequences, with which components.

\subsection{Final remark}
We have insisted on numerical examples, but the co-recursive strategy is, of course, domain independent, provided that an adequate set of mathematical operation is defined. The last example combined two different sorts of sequences.

It is possible to write such algorithms in a symbolic setting, and we did it in Maple and in MuPAD\@. But the language of these packages is strict; we used unevaluated macros in order to simulate the lazy processing, and the ``intermediate expression swell disease'' is a serious hindrance. Probably the languages based on rewriting, such as Mathematica, are better adapted to this kind of manipulations than the very procedural Maple, but we haven't tested it.


\section{Exercise: Feynman diagrams in a 0-dimensional quantum field theory}
Your task is to generate the full, untruncated perturbational expansion for a toy Quantum Field Theory, which describes the scattering and producing of scalar, chargeless quantum particles in a zero-dimensional space-time. You will generate infinite power series in the coupling constant $\gamma$, considered small. Despite the lack of realism, this is a genuine computational problem, used sometimes to generate combinatorial factors for a more serious model. You might appreciate that three pages of description reduce finally to a program, whose essential part is a one-liner! 

In Quantum Field Theory {\bf nothing} is easy. However, a substantial part of the computational difficulties comes from the fact that the only effective mechanism we have, the perturbational calculus, produces {\itbf open, extrapolating} recurrence equations for some expressions, and the classical coding of the higher-order terms is incredibly intricate. This can be seen even in a toy theory, where there is no space, and all ``interactions'' and ``propagation'' of quantum particles, reduce to pure numbers~\cite{PREDRAG}.
\begin{figure}[H]
\centering
 \vspace{-0.3cm}
 \includegraphics[width=13cm,keepaspectratio]{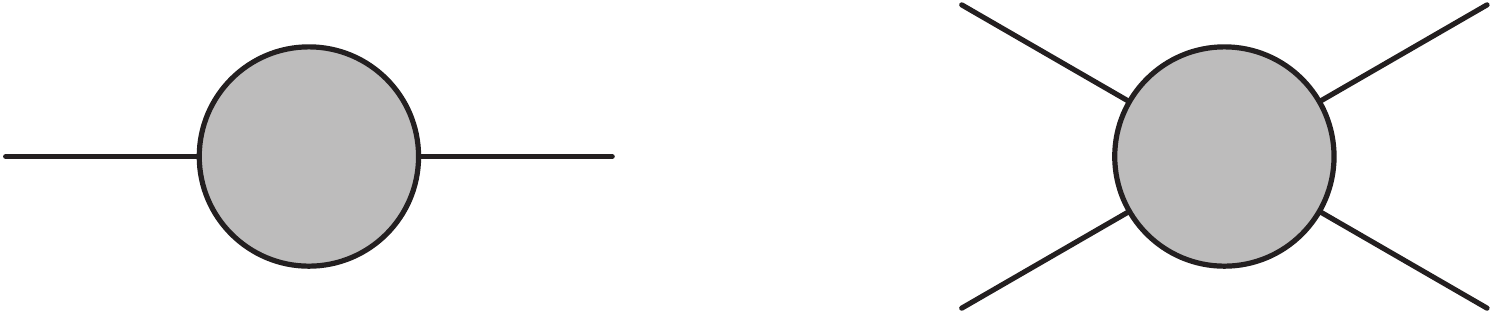}
 \vspace{-0.40cm}
 \caption{General behaviour of particles, propagation and interaction}
 \vspace{-0.40cm}
 \label{gfunc}
\end{figure}
\noindent
The theory possesses some objects depicted above. The first is the ``propagator'', the probability amplitude (you don't need to know its true meaning) that a particle moves from one place to another. In a true field theory, this is a function $G_2(x_0;x_1)$ depending on the space coordinates, but here it will be just a number $G_2$. The second is the amplitude of scattering, $G_4$, which in a realistic theory is $G_4(x_0,y_0;x_1,y_1)$. The series for $G_2$ and $G_4$, which we include for comparison with your results, is:
$$
G_2 = 1 + \gamma^2 + \frac{25}{8}\gamma^4 + 15\gamma^6 + \frac{12155}{128}\gamma^8 + \frac{11865}{16}\gamma^{10} + \frac{7040125}{1024}\gamma^{12} + \cdots
$$
$$
G_4 = \frac{1}{2}\gamma^2 + 4\gamma^4 + \frac{525}{16} \gamma^6 + 300 \gamma^8 + \cdots
$$
\begin{wrapfigure}[9]{r}{3.2cm}
  \centering
 \vspace{-0.80cm}
 \includegraphics[width=3.2cm,keepaspectratio]{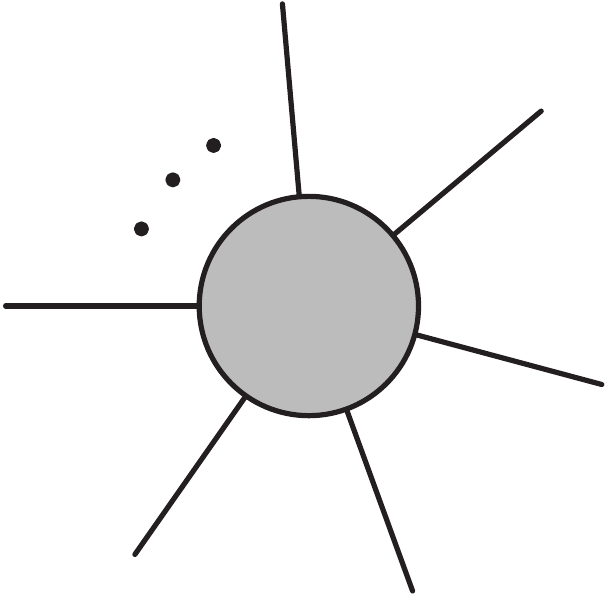}
 \vspace{-0.90cm}
\caption{Arbitrary interaction}
\end{wrapfigure}
(By the way, the cited above book of Cvitanovi\v{c} contains a mistake in one of the computed coefficients. We are sure that our solution is correct, since we had simply no place to commit errors!) Our formalism should compute the amplitudes for any number of particles present, and their repartition in time, e.g.~with two particles coming in, and six getting out, since they can be created or annihilated. (Pathological cases, such as: some particles entering, and nothing coming out, or {\em vice-versa}, should be excluded, since this is physics, and not the theory of finance\dots). Readers not interested in physics may consider this exercise as a Graph Theory problem.

The theoretical basis is poor, we have just two primitive elements, the ``bare'' propagator $\Delta$, which is a given function $\Delta(x_0;x_1)$, and describes the motion of a particle without any interaction. Here it is a number, and it can be attributed any value, for example 1. The second is the bare vertex $V$, describing an emission (or an absorption, this is absolutely equivalent, since the model is timeless) of a virtual particle. Their graphical form is straightforward:
\begin{figure}[H]
\centering
\vspace{-0.30cm}
 \includegraphics[width=7cm,keepaspectratio]{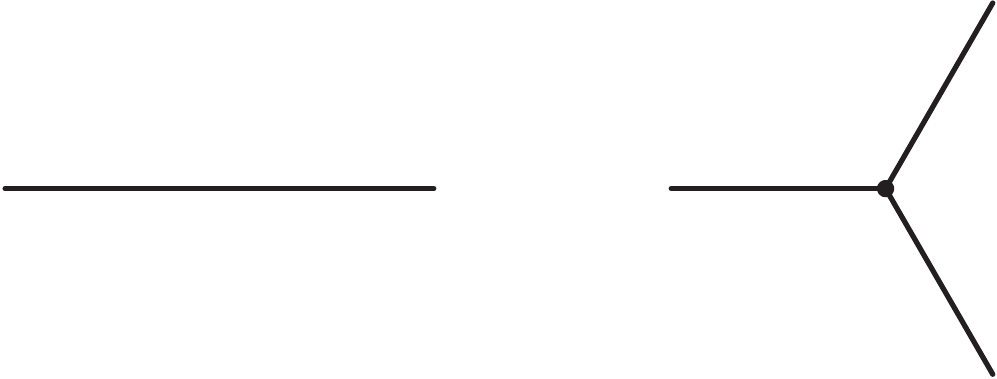}
 \vspace{-0.60cm}
 \caption{Primitive propagator and vertex}
 \label{primi}
\end{figure}
\noindent
The vertex diagram may also represent the disintegration of a particle in two, or the reverse process. The primitive vertex is numerically proportional to a {\em small} number $\gamma$~-- the coupling constant, and in our model is just equal to it. Attention, in fact the drawing at the right may be considered as the vertex attached to three propagators: $\gamma \Delta^3$, the vertex itself is just the ``dot'', or the dot with ``hooks'': \ \includegraphics[width=0.3cm,keepaspectratio]{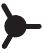}. 

The interaction is ``costly'', and {\bf the amplitudes will be represented as power series in $\gamma$}. The first terms should give a reasonable approximation, although in known interesting models these series are asymptotic, and diverge. But now {\em everything} may happen. The theory says that the amplitude for a given process is equal to the sum of the weights of {\em all possible} contributing diagrams, e.g.:
\begin{figure}[H]
\centering
\vspace{-0.20cm}
 \includegraphics[width=16cm,keepaspectratio]{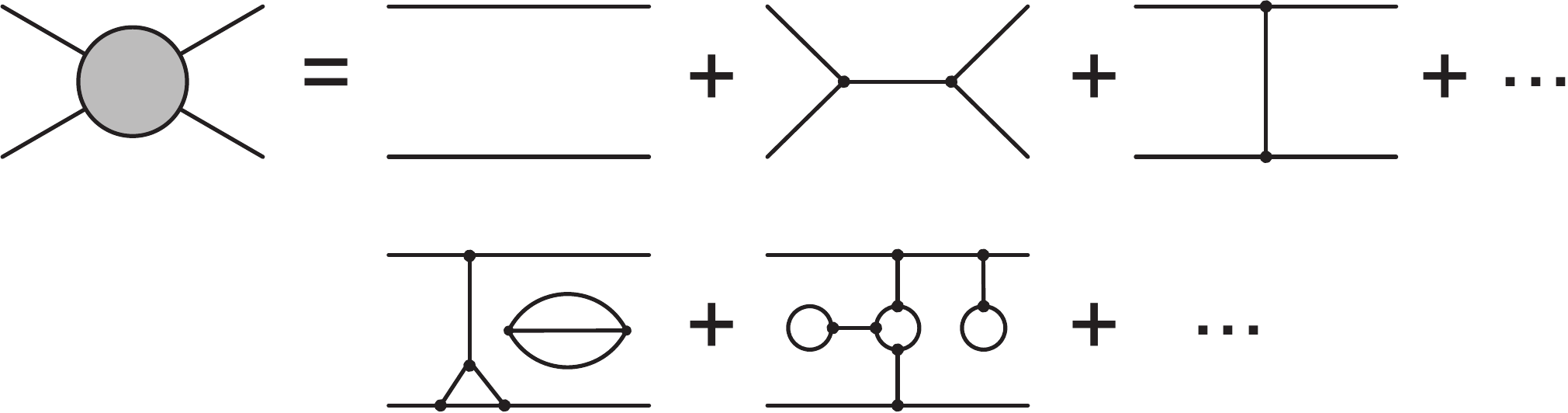}
 \vspace{-0.60cm}
 \caption{Some diagrams\ldots}
 \label{alldia}
\end{figure}
\noindent
{\bf Exercise.} Draw {\em all the diagrams} for $G_3$ with 3 internal vertices. Write the analytical forms (using $\Delta$ and $\gamma$) for their weights. {\bf Question.} Is it obvious that in $G_{2n+(0,1)}$ only the terms with $\gamma^{2m+(0,1)}$ are present? Why? (You don't need to be a Field Theory specialist, to answer it.)

\mnd
We have to sum all {\em topologically different} diagrams, and the idea is to generate them recursively. We will ``attach'' to the external edges some fictitious variables $J$ called {\em currents} (depicted as crosses) $J$, in order to construct the equation fulfilled by the generating functional
\begin{equation}
Z[J] = \sum_{n}^\infty \frac{1}{n!} G_n J^n\,,
\end{equation}
\begin{figure}[H]
\centering
\vspace{-0.20cm}
\includegraphics[width=6cm,keepaspectratio]{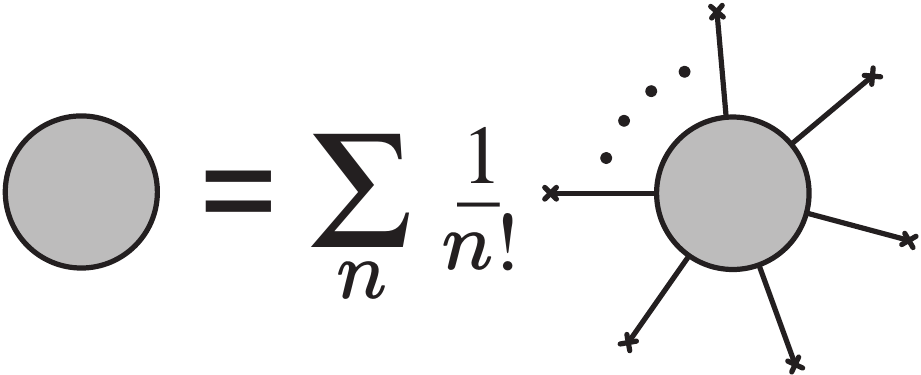}
\vspace{-0.20cm}
\caption{Generating functional}
\end{figure}
\noindent
(In a realistic theory $J$ are {\em functions}, and the functional involves the integration over the coordinates.) The sum above is equivalent to $G_n = \frac{d^n}{d J^n} Z[J]|_{J=0}$. One differentiation produces one external leg. 

Our theory must be complete (this is the meaning of: ``everything may happen''). A particle may pass through the interaction zone without interaction, or it may scatter at least once. It is easy to see that the interaction diagram for say, $G_4$ may be drawn as
\begin{figure}[H]
\centering
\vspace{-0.20cm}
 \includegraphics[width=16cm,keepaspectratio]{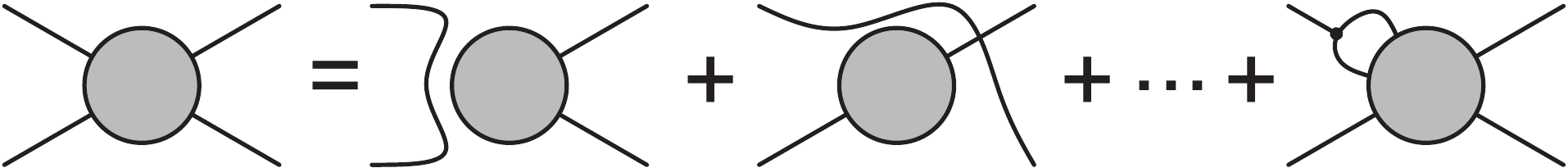}
 \vspace{-0.60cm}
 \caption{Recursive reduction of $G_4$}
 \label{recgfour}
\end{figure}
\noindent
Notice that $G_4$ contains $G_5$! The formalism is naturally co-recursive (although rare are physicists who have ever heard this name\ldots). We shall repeat the same reduction in the case of $Z[J]$. From now on the ``bubbles'' will implicitly keep the dependence on $J$, we will not put implicitly $J=0$.
\begin{figure}[H]
\centering
\vspace{-0.20cm}
 \includegraphics[width=12cm,keepaspectratio]{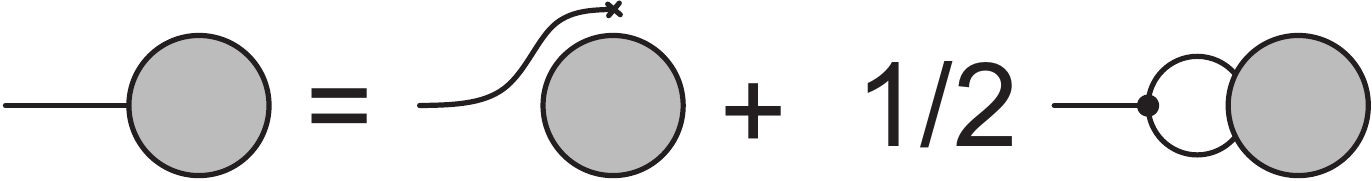}
 \vspace{-0.20cm}
 \caption{Dyson-Schwinger equation}
 \label{dyson}
\end{figure}
\noindent
{\bf Exercise.} Prove that its analytical form is:
\begin{equation}
\frac{d}{d J} Z[J] = J \Delta \cdot Z[J] + \frac{1}{2} \Delta \gamma \frac{d^2}{d J^2} Z[J]
\end{equation}
Is it obvious that the factor $1/2$ which comes from the symmetry considerations is necessary?

\mnd
{\bf Exercise.} In the theory of graphs we know that if $Z$ is the generating functional for a set of graphs, its logarithm: $W = \log Z$ generates all the connected components. Write the D-S equation for the first derivative of $W$, and draw the appropriate diagram. Prove that the diagram is:
\begin{figure}[H]
\centering
\vspace{-0.20cm}
 \includegraphics[width=16cm,keepaspectratio]{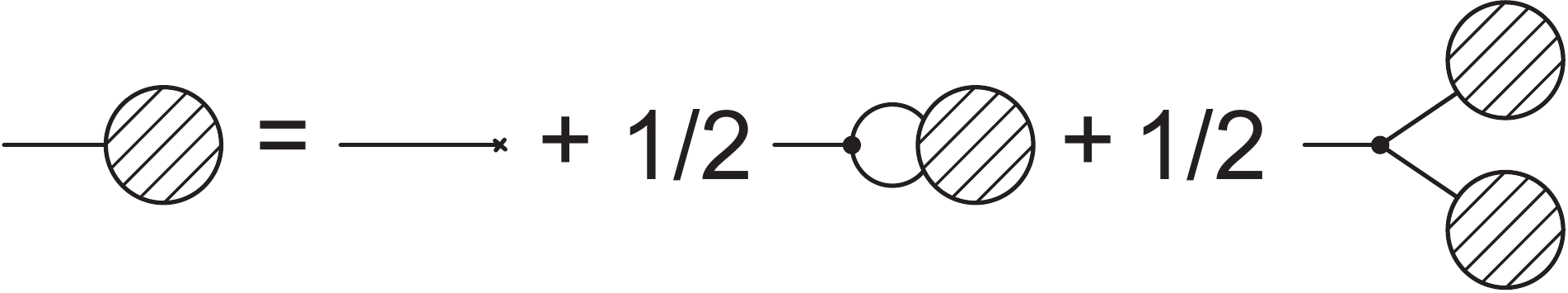}
 \vspace{-0.60cm}
 \caption{Logarithmic Dyson-Schwinger equation}
 \label{dyslog}
\end{figure}
\noindent
The combinatorics of this form is similar to the D-S equation for a real theory, but the algebra is richer (spinors, charge indices, etc., and the space-time adds many analytic features, the functional (or at least partial) derivatives, and the space integration. In principle the development of $W$ in powers of $\gamma$ may, and has been used to generate approximations to measurable amplitudes in electrodynamics, but using some other techniques, not the D-S equation. Its co-recursive form makes it difficult to generate the code. But not for us!

$W$ is a function of $J$ and $\gamma$, a double series, which should be properly ordered: we shall need some first terms of the development in $J$ ($G_2$, or $G_4$, etc.), but we want the full, not truncated series in $\gamma$. 

{\bf Exercise.} Introduce $\varphi = dW/dJ$, and rewrite the equation of the Fig. \ref{dyslog}, assuming that $\Delta=1$; this parameter, related to the ``mass'' of the particle may be arbitrarily fixed. Prove that you get: $\varphi = J +\frac{1}{2}\gamma(\varphi' + \varphi^2)$. Write a program which solves this equation and computes $\varphi$. Notice that there is almost nothing to do, this equation {\em should be} the program. But the representation of terms depends on the ordering of the series. Consider $\varphi$ first as a series in $\gamma$, whose coefficients are series in $J$, the first element of the result, $\varphi_0(J) = J$ is the identity series. The differentiation $\varphi'$ propagates to the coefficients, so it should be coded as {\tt fmap sdiff}. 

{\bf Exercise.} Compute $G_2$ by taking the second term of each element of $\varphi$, and $G_4$~-- the fourth term. Or, better: {\itbf transpose the series}, and take its second and the fourth element. The transposition function should be coded in one line\ldots

\section{Final remarks}
Lazy programming is sufficiently universal so that we may speak about a specific language, which has to be mastered, with its semantics, and style. This style is not a free lunch\ldots{} Some ``simple'' algorithms have fast rising complexity, and the programs do horrible things with the memory. (In some cases a memoization may help). Sometimes a serious human preprocessing of the formulae is needed in order to yield a sound co-recursive algorithm. However, we are persuaded that the positive aspects of this approach dominate.

While discussing the acoustic streams we mentioned that often we don't need to generate the ``time'' stream. But it may be needed elsewhere, e.g.~in the solution of equations of motion. The arithmetic sequences: {\tt [x0, x0+h ..]} are so ubiquitous and useful, that it might be interesting to represent them just through their first element, and the difference {\tt h}. The presentation above may and should be optimized~\cite{LASUI}.

\subsection{Co-recursion, Universe and Everything}
Since this is a ``distilled tutorial'', perhaps a bit of a ``distilled philosophy'' (ethereal\ldots) might be appropriate. In a computing world where imperative programs dominate, the lazy, co-recursive data are unknown, and it is easy to claim that this approach is ``unnatural''. However, the {\em extrapolation}, and even the extrapolating self-reference is quite popular. The well-known flood-filling algorithm in computer graphics is a paradigmatic example thereof. In order to fill a figure (contiguous set of pixels) with a given colour, select one (test if unfilled) pixel, fill it, and repeat the same for all the neighbours. This is the construction of the transitive closure of the neighbouring relation, and it is also used in the simulation of the percolation, and other similar phenomena. The process is ``potentially infinite'' and only the finiteness of the space ensures its termination. But even if the space is infinite, the {\em finite progress} of the process~-- essential for the co-recursion~-- is a sound property. The process may be interrupted at any moment, and resumed.

Why $s_{n+1} = f(s_n)$ (given $s_0$) should be more natural than $s = s_0 :> f(s)$ (with an appropriate sequencing constructor $(:>)$ which replaces the assignment of some local variables)? Is the strictness, the applicative order of evaluation more natural than the laziness? The deferred execution {\em might} be slightly less efficient, but we gain in modularity, the creation and the consumption of data are conceptually separate (although physically they are interleaved). The main obstacle for the usage of co-recursive techniques is the teaching habitude, the mental pattern ``borrow from the future'' seems a bit exotic.

But is it? In my opinion a small child who learns to walk, applies this pattern. She has no ``administrative memory'' in order to construct an iteration loop in her head. The walking process consists in traversing a ``co-recursive sequence'' of steps, while adding them to this Great Plan. And everybody who had children knows how this process stops\ldots{}

The next example may look a little cynical, but it is realistic. Borrowing money (bank credits) by poor countries (and some families in difficulty) is too often such a process: more money is borrowed and added to the debt, in order to pay the existing credit (to traverse the sequence of reimbursement steps). We know~-- unfortunately~-- that this infinite co-recursion must be often broken by an external mechanism\ldots{} The very idea of the bank credit is a lazy, co-recursive (runaway) algorithm, which works in practice because the bank buffers are voluminous enough. The true Game consists in running away before the Bottom hits you. By the way, in some countries (we shall omit their names) the social security expenses operate on borrowed-from-the-future funds, which will (hopefully) ``un-virtualize'' themselves thanks to the work of our grandchildren. 

If the reader is still unconvinced, here is the Ultimate Example showing that the co-recursion is The Big Thing. We have seen that the perturbational approach in the Quantum Field Theory demands that the description of the virtual amplitudes with $n$ particles contain the amplitudes with $n+1$, $n+2$, etc.\ bodies. But, is it just a particular theoretical description? No, we {\em know} that the theory works, that in the electro-weak interactions the particles get produced according to this scheme, and that they un-virtualize and become real, provided there is enough energy around. But now, in Cosmology, there is a (metaphysical; impossible to prove, but {\em please! have faith!}) hypothesis that the negative gravitational potential energy of all particles, compensates exactly their positive energy due to their masses. So, the virtual process of creating more, and more particles, and adding them to this co-recursive Universe may borrow (let's be honest: steal!) the energy from the gravitation, activated when the particles will appear\ldots{} The quantum fluctuations become persistent, and here we are. There cannot be any doubt: ``{\tt let} there be light!'' is a lazy (co-recursive) functional construct.

Now, perhaps the Universe will pay the debt one day, or perhaps the Processor decides to raise a particular exception to stop this silly process, but this is a matter for another tutorial, presented when I learn the answer.

\bibliographystyle{eptcs}
\bibliography{scidsl}
\end{document}